\documentclass{ws-procs9x6}

\newcommand {\bd}{\begin{displaymath}}
\newcommand {\ed}{\end{displaymath}}
\newcommand {\eq}{\begin{equation}}
\newcommand {\beq}{\begin{equation}}
\newcommand {\eeq}{\end{equation}}
\newcommand {\beqa}{\begin{eqnarray}}
\newcommand {\eeqa}{\end{eqnarray}}

\newcommand {\tr}{{\rm tr\,}}

\newcommand {\RR}{\mbox{\scriptsize R}}
\newcommand {\II}{\mbox{\scriptsize I}}

\newcommand {\trs}{\mbox{\scriptsize tr}}

\newcommand {\ee}{\mbox{e}}

\newcommand {\dd}{\mbox{d}}

\newcommand {\defeq}{\stackrel{\rm def}{=}}

\newcommand {\vev} [1] {\langle #1 \rangle}
\newcommand{\id}{{1\!\!1}} 

\renewcommand{\theequation}{\thesection.\arabic{equation}}

\begin{document}

\title{Factorization Method for Simulating\\
QCD at Finite Density}

\author{Jun NISHIMURA\footnote{\uppercase{W}ork partially
supported by \uppercase{G}rant-in-\uppercase{A}id for 
\uppercase{S}cientific \uppercase{R}esearch (\uppercase{N}o.\
14740163) 
from the \uppercase{M}inistry of \uppercase{E}ducation, 
\uppercase{C}ulture, \uppercase{S}ports, \uppercase{S}cience and 
\uppercase{T}echnology.}}

\address{Department of Physics, Nagoya University, \\
Nagoya 464-8602, Japan\\
E-mail: nisimura@eken.phys.nagoya-u.ac.jp}


\maketitle

\abstracts{
We propose a new method for simulating QCD at finite density.
The method is based on a general factorization property of 
distribution functions of observables,
and it is therefore applicable to {\em any} system with a complex
action. The so-called overlap problem is completely eliminated
by the use of constrained simulations.
We test this method in a Random Matrix Theory for finite density
QCD, where we are able to reproduce the exact results for the 
quark number density. 
}

\section{Introduction}
Recently there are a lot of activities in QCD 
at finite density, 
where interesting phases such as a superconducting phase
have been conjectured to appear \cite{bailin}.
At zero chemical potential
Monte Carlo simulations of lattice QCD
enables nonperturbative studies from first principles.
It is clearly desirable to extend such an approach 
to finite density and explore the phase diagram of QCD in the
$T$(temperature)-$\mu$(chemical potential) plane.
The main obstacle is here that the Euclidean action becomes complex
once the chemical potential is switched on.

Nevertheless QCD at finite density has been studied by various
approaches with exciting conjectures.
First there are perturbative studies which are valid 
in the $\mu \rightarrow \infty$ limit \cite{Son:1998uk,Schafer:1999jg}.
Refs.\ [\refcite{krishna}] and [\refcite{edward}]
uses effective theories
with instanton-induced four-fermi interactions.
As for Monte Carlo studies two directions have been pursued so
far. One is to modify the model so that the action becomes
real. This includes changing the gauge group \cite{Kogut:2002kj}
from SU(3) to SU(2),
and introducing a chemical potential with opposite signs for up and down
quarks \cite{Kogut:2002se}.
%
The other direction is to explore the large $T$ and small
$\mu$ regime of lattice QCD, where the imaginary part of the action is
not very large \cite{Fodor:2001au,Allton:2002zi}. 
These studies already produced results relevant to heavy ion
collision experiments,
but more interesting physics will be uncovered 
if larger $\mu$ regime becomes accessible by simulations.

In Ref.\ [\refcite{Anagnostopoulos:2001yb}]
we have proposed 
a new method to simulate systems with a complex action,
which utilizes a simple factorization property of 
distribution functions of observables.
Since the property holds quite generally,
the approach can be applied to {\em any} system with a complex action.
The most important virtue of the method is that
it eliminates the so-called {\em overlap
problem}, which occurs in the standard re-weighting method.
Ultimately we hope that this method will enable us, among other things,
to explore the phase diagram of QCD at finite baryon density.
As a first step we have tested 
\cite{Ambjorn:2002pz}
the new approach in a Random Matrix Theory for finite density 
QCD \cite{Stephanov:1996ki}, which can be regarded as a schematic
model for QCD at finite baryon density.

\section{Random Matrix Theory for finite density QCD}
\label{RMTintro}
\setcounter{equation}{0}
\renewcommand{\thefootnote}{\arabic{footnote}}

The Random Matrix Model we study is defined by the 
partition function
\beq
Z = \int \dd W \ee ^{- N \, \trs (W^\dag W)} \, \det D  \ ,
\label{rmtdef}
\eeq
where $W$ is a $N \times N$ complex matrix, and 
$D$ is a $2N \times 2N$ matrix given by
\beq
D = 
\left( 
\begin{array}{cc}
m & i W +  \mu  \\
i W^\dag  + \mu & m
\end{array}
\right) \ .
\label{defD}
\eeq
The parameters $m$ and $\mu$ correspond to the 
`quark mass' and the `chemical potential', respectively. 
In what follows we consider the massless case ($m=0$) for simplicity
and we focus on the `quark number density' defined by
\beq
\nu   =  \frac{1}{2N} \, \tr \, ( \gamma_4 D^{-1} ) \ ,
\mbox{~~~~~~~} \gamma_4 = \left( 
\begin{array}{cc}
0 & \id  \\
\id  & 0
\end{array}
\right) \ .
\eeq
The vacuum expectation value (VEV) of the quark number density is
obtained exactly by [\refcite{Halasz:1997he}]
and in particular in the large-$N$ limit \cite{Stephanov:1996ki}
\beq
\lim_{N\rightarrow \infty}
 \langle \nu \rangle =
\left\{ \begin{array}{ll}
- \mu  & \mbox{for~}\mu < \mu _{\rm c}  \\
1/ \mu  & \mbox{for~}\mu > \mu _{\rm c} \ ,
\end{array} 
\right. 
\label{nq_rmt}
\eeq
where $\mu_{\rm c}$ is the solution to
the equation $1 + \mu^2 + \ln (\mu^2) = 0$,
and its numerical value is given by $\mu_{\rm c} =0.527\cdots$.
We find that the quark number density $\langle \nu \rangle$
has a discontinuity at $\mu = \mu_{\rm c}$.
Thus the schematic model reproduces qualitatively 
the first order phase transition expected to occur
in `real' QCD at nonzero baryon density.

\section{The complex action problem}
\label{problem}
\setcounter{equation}{0}
\renewcommand{\thefootnote}{\arabic{footnote}}

Let us first rewrite (\ref{rmtdef}) as
\beq
Z = \int \dd W \, \ee ^{-S_0 + i \, \Gamma}\ ,
\label{rmtdef2}
\eeq
where we have introduced $S_0$ and $\Gamma$ by
\beqa
S_0 &=& N \, \tr (W ^\dag W) - \ln | \det D | \\
\det D &=&  \ee ^ {i \Gamma}\,  | \det D | \ .
\eeqa
In this form it becomes manifest that the system has a 
{\em complex action}, where the
problematic imaginary part $\Gamma$
is given by the phase of the fermion determinant.
Since the weight $\ee ^{-S_0 + i \, \Gamma}$ in (\ref{rmtdef2})
is not positive definite, 
we cannot regard it as a probability density.
Hence it seems difficult to apply the idea of standard Monte Carlo
simulations,
which reduces the problem of obtaining VEVs to that of taking 
an average over an ensemble generated by the probability density.

Let us define the so-called {\em phase quenched} partition function
\beq
Z_0 = \int \dd W \, \ee ^{- N \, \trs (W^\dag W)} \, | \det D | 
= \int \dd W \, \ee ^{-S_0}\ .
\label{absdef}
\eeq
Since the system (\ref{absdef}) has a positive definite weight,
the VEV $\langle \ \cdot \ \rangle _{0}$ associated with
this partition function can be evaluated by
standard Monte Carlo simulations.
Then one can use the standard re-weighting formula
\beq
\left\langle \nu \right\rangle
= \frac{\left\langle \nu \, \ee ^{i \Gamma }
\right\rangle _{0}}
{\left\langle \ee ^{i \Gamma }
\right\rangle _{0}}    
\label{VEV}
\eeq
to obtain the VEV $\langle \nu \rangle$ in the full model
(\ref{rmtdef2}).
The problem with this method is that 
the fluctuations of the phase $\Gamma $ in (\ref{VEV}) 
grows linearly with the size of the matrix $D$,
which is of O($N$).
Due to huge cancellations,
both the denominator and the numerator of the r.h.s.\ of (\ref{VEV})
vanish as $\ee ^{-{\rm const.} N}$ as $N$ increases,
while the `observables' 
$\ee ^{i \Gamma }$ and $\nu \ee ^{i \Gamma }$
are of O(1) for each configuration.
As a result, 
the number of configurations required to obtain the VEVs with
some fixed accuracy grows as $\ee ^{{\rm const.} N}$.

In fact we may simplify the expression (\ref{VEV}) slightly
by using a symmetry.
We note that the fermion determinant $\det D$,
as well as the observable $\nu$,
becomes complex conjugate under the transformation
\beq
W \mapsto -  W \ , 
\label{transf}
\eeq
while the Gaussian action remains invariant.
From this we find that
\beqa
\label{sum_nu}
&~& \langle \nu \rangle = \langle \nu_{\rm R} \rangle
+ i \, \langle \nu_{\rm I} \rangle  \\
&~&  \langle \nu_{\rm R} \rangle =
\frac{\left\langle \nu_{\rm R} \cos \Gamma 
\right\rangle _{0}}
{\left\langle \cos \Gamma 
\right\rangle _{0}}  ~~~;~~~
\langle \nu_{\rm I} \rangle =
i \, \frac{\left\langle \nu_{\rm I} \sin \Gamma 
\right\rangle _{0}}
{\left\langle \cos \Gamma 
\right\rangle _{0}}  \ ,
\label{reweight}
\eeqa
where $\nu_{\rm R}$ and $\nu_{\rm I}$ denote the real part
and the imaginary part of $\nu$, respectively.
This simplification, however, does not solve the problem at all,
since $\, \cos  \Gamma \, $ and $\, \sin  \Gamma \, $ 
flip their sign violently as a function of the configuration $W$.
Note that both terms in the r.h.s.\ of (\ref{sum_nu}) are real,
meaning in particular that their sum $\langle \nu \rangle$ is also
real.

The model (\ref{absdef}) is solvable in the 
large-$N$ limit \cite{Stephanov:1996ki} and one obtains
\beq
\lim_{N\rightarrow \infty}
\langle \nu \rangle_0 = 
\left\{ \begin{array}{ll}
\mu  & \mbox{for~}\mu < 1  \\
1/ \mu   & \mbox{for~}\mu > 1 \ .
\end{array} 
\right. 
\label{nq_abs}
\eeq
In this case the VEV of the quark number density is a continuous function
of the chemical potential $\mu$ unlike in (\ref{nq_rmt}).
Thus the first order phase transition
in the full model (\ref{rmtdef2})
occurs precisely due to the imaginary part $\Gamma$ of the action.
Note also that the symmetry under (\ref{transf}) implies
\beq
\langle \nu_{\rm I}\rangle _0 = 0 ~~~;~~~
\langle \nu _{\rm R} \rangle _0 =\langle \nu \rangle _0 \ .
\label{RI0}
\eeq

\section{The factorization method}
\label{method}
\setcounter{equation}{0}
\renewcommand{\thefootnote}{\arabic{footnote}}


In this section, we explain how the factorization method 
\cite{Anagnostopoulos:2001yb}
can be used to obtain the VEVs $\langle \nu_{\rm R} \rangle$
and $\langle \nu_{\rm I} \rangle$.
The fundamental objects of the method
are the distribution functions
\beqa
\rho_i(x) &\defeq& \langle \delta (x - \nu_i  )\rangle \\
\rho^{(0)}_i (x) &\defeq& \langle \delta (x - \nu_i )\rangle_0
~~~~~~i={\rm R,I} 
\eeqa
defined for the full model and for the phase quenched model respectively.
The important property of these functions is that they factorize as
\beq
\rho_i(x) = \frac{1}{C} \, \rho^{(0)}_i (x) \, \varphi _i(x)  
~~~~~~i={\rm R,\,I} \ ,
\label{factorize}
\eeq
where the constant  $C$ is given by
$C \defeq \langle \ee ^{i \, \Gamma}\rangle _0$.
The `weight factor' $\varphi _i(x)$ represents the effect of 
$\Gamma$, and it can be written as a VEV
\beq
\varphi_i(x) \defeq \langle \ee ^{i\, \Gamma} \rangle_{i,\, x}
\eeq
with respect to a yet another partition function
\beq
Z_i(x) = \int \dd W \, \ee^{-S_0} \, \delta(x-\nu_i) \ .
\label{cnstr_part}
\eeq
The $\delta$-function represents a constraint on the system.
In actual simulation we replace 
the $\delta$-function by a sharply peaked potential.
We refer the reader to Ref.\ [\refcite{Ambjorn:2002pz}]
for the details.


Using the symmetry under (\ref{transf}), the formulae 
for $\vev{\nu}$ is nothing but (\ref{reweight}), where
$\left\langle \nu_{\rm R} \cos \Gamma 
\right\rangle _{0}$,
$\left\langle \nu_{\rm I} \sin \Gamma 
\right\rangle _{0}$
and $\left\langle \cos \Gamma \right\rangle _{0}$ 
are replaced by
\beqa
\label{simpleformula0}
\left\langle \nu_{\rm R} \cos \Gamma 
\right\rangle _{0}  
&=& \int _{-\infty} ^{\infty}
 \dd x \, x \, \rho_{\rm R}^{(0)} (x) \, w_{\rm R}(x) \ , \\
\left\langle \nu_{\rm I} \sin \Gamma 
\right\rangle _{0}
&=&
2 \int  _{0} ^{\infty}
\dd x \, x \, \rho ^{(0)}_{\rm I} (x) \, w_{\rm I} (x)
\label{simpleformula} \ ,\\
\left\langle \cos \Gamma 
\right\rangle _{0}
&=& 
\int _{-\infty} ^{\infty} \dd x 
\, \rho^{(0)} _{\rm R} (x) \,  w _{\rm R} (x) \ .
\label{C_new}
\eeqa
The weight factors $w_i(x)$ are defined by
\beq
\label{wRcos}
w_{\rm R}  (x) \defeq \langle  \cos \Gamma \rangle_{\RR , x }  
\mbox{~~~};\mbox{~~~}
w_{\rm I}(x) \defeq \langle \sin \Gamma\rangle_{\II , x}  \ .
\eeq
%
%

One of the virtues of the method can be seen
from (\ref{simpleformula0})$\sim$(\ref{C_new}).
If we are to obtain the VEVs on the l.h.s.\
by directly simulating the system (\ref{absdef}),
for most of the time we sample configurations
whose $\nu_i$ takes a value close to the peak
of $\rho_i^{(0)} (x)$.
However, from the r.h.s.\ of the formulae,
it is clear that 
we have to sample configurations whose $\nu_i$ takes a value 
where $\rho_i^{(0)} (x) |w_i(x)|$ 
becomes large, in order to obtain the VEVs accurately.
In general these two regions of configuration space
have little overlap, which becomes exponentially small as the system
size increases.
The present method resolves this `overlap problem' completely
by `forcing' the simulation to sample the important region.

\begin{table}[ph]
\tbl{Results of the analysis of $\langle \nu \rangle $
described in the text.
Statistical errors computed by the jackknife method are shown.
The last column represents the exact result 
for $\vev{\nu}$ at each $\mu$ and $N$.
For $\mu=0.2$ the exact result is $\vev{\nu}=-0.2$ 
with an accuracy better than $1$ part in $10^{-9}$. }
{\footnotesize
\begin{tabular}{|c | c | c | c | c | c | }
\hline
$\mu$&$N$ &$\vev{\nu_{\rm R}}$&$i \, \vev{\nu_{\rm I}}
$&$\vev{\nu}$&$\vev{\nu}$~(exact)\\
\hline
 0.2 & 8  & 0.0056(6) & -0.1970(5)  & -0.1915(7) & -0.20000\ldots \\
 0.2 & 16 & 0.0060(4) & -0.1905(13) & -0.1845(13)& -0.20000\ldots \\
 0.2 & 24 & 0.0076(9) & -0.1972(14) & -0.1896(17)& -0.20000\ldots  \\
 0.2 & 32 & 0.0021(8) & -0.1947(19) & -0.1927(25)& -0.20000\ldots  \\
 0.2 & 48 & 0.0086(37)& -0.2086(54) & -0.2000(88)& -0.20000\ldots  \\
\hline
 1.0 &  8 & 0.8617(10)&  0.1981(13) &  1.0598(12)&1.066501$\ldots$\\
 1.0 & 16 & 0.8936(2) &  0.1353(6)  &  1.0289(5) &1.032240$\ldots$\\
 1.0 & 32 & 0.9207(1) &  0.0945(2)  &  1.0152(3) &1.015871$\ldots$ \\
\hline
\end{tabular}
\label{t:1} }
\end{table}

\section{Reproducing exact results by the new method}
\label{results}
\setcounter{equation}{0}

In Table\ \ref{t:1} we show our results for two values of $\mu$, 
$\mu = 0.2$ and $\mu = 1.0$, which are on opposite  
sides of the first order 
phase transition point $\mu = \mu_{\rm c}=0.527 \cdots$.
They are in good agreement with the exact results,
and the achieved values of $N$ are 
large enough to extract the large $N$ limit.

Note that $\langle \nu_{\rm R} \rangle \sim 0$ at $\mu = 0.2$.
Thus the main contribution to $\langle \nu \rangle$ comes from the 
imaginary part $\langle \nu_{\rm I} \rangle$,
which is in sharp contrast to 
the results (\ref{RI0}) for the phase quenched system.
This result comes about because the sign change of $w_{\rm R}(x)$
occurs near the peak of $\rho _{\rm R} ^{(0)} (x)$,
so that the product 
$\rho _{\rm R} ^{(0)} (x)w_{\rm R}(x)$ 
has a positive regime and a negative regime, which cancel
each other in (\ref{simpleformula0}).
For $\mu = 1.0$, on the other hand,
$w_{\rm R}(x)$ is approximately constant
in the region where $\rho _{\rm R} ^{(0)} (x)$ is peaked,
so the shape of the product 
$\rho _{\rm R} ^{(0)} (x)w_{\rm R}(x)$ 
is similar to $\rho _{\rm R} ^{(0)} (x)$.
The main contribution to $\langle \nu \rangle$ comes from the 
real part $\langle \nu_{\rm R} \rangle$,
and moreover, it is close to $\langle \nu_{\rm R} \rangle_0$.

\begin{figure}[htbp]
  \begin{center}
    \includegraphics[height=7cm]{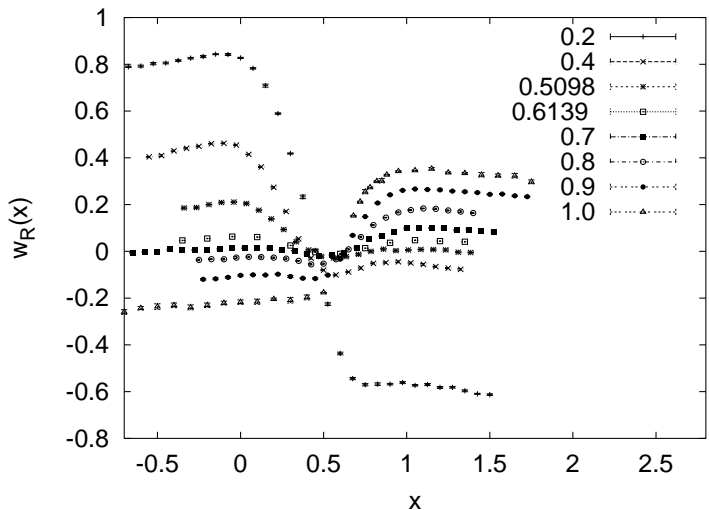}
    \caption{The weight factor $w_{\rm R}(x)$ is plotted 
against $x$ 
for $N=8$ at various $\mu$.
The behavior changes drastically as $\mu$ crosses 
the critical point.}
    \label{fig:wR_N8}
  \end{center}
\end{figure}

In Fig.\ \ref{fig:wR_N8}
we plot $w_{\rm R}(x)$
for $N=8$ at various $\mu$.
It is interesting that
the $w_{\rm R}(x)$ changes from positive to negative for 
$\mu < \mu_{\rm c}$,
but it changes from negative to positive for $\mu > \mu_{\rm c}$.
(Similarly $w_{\rm I}(x)$ is positive at $x>0$ for $\mu < \mu_{\rm c}$,
but it is negative at $x>0$ for $\mu > \mu_{\rm c}$.)
Thus the behavior of $w_i (x)$ changes drastically as 
the chemical potential $\mu$ crosses its critical value $\mu_{\rm c}$.
These results provide a clear understanding of 
how the first order phase transition occurs
due to the effects of $\Gamma$.



\section{Applications to other systems with complex actions}
\label{Summary}
\setcounter{equation}{0}

The method \cite{Azcoiti:2002vk}
proposed for simulating $\theta$-vacuum like systems
can be regarded as a {\em special} case of the factorization method.
A simplified version of the method was sufficient because
the observable was identical to the imaginary part of the action.
The essence of the factorization method is that it avoids the
overlap problem by the use of constrained 
simulations.
In Ref.\ [\refcite{Azcoiti:2002vk}]
promising results for 2d CP$^3$ are also reported.

In Ref.\ [\refcite{Anagnostopoulos:2001yb}]
the method has been used
to study the dynamical generation
of space time in superstring theory based on its matrix model
formulation \cite{IKKT}.
There the method becomes even more
powerful since the distribution functions
turn out to be positive definite.
In this case
the scaling property of the weight factor
enables extrapolations to larger system size.

We hope that the factorization method is useful also for
studying other interesting systems with complex actions such as
Chern-Simons theories, chiral gauge theories, 
strongly coupled electron systems etc.



\end{document}